\documentclass[prb,preprint,showpacs]{revtex4}
\usepackage{revsymb}
\usepackage[dvips]{graphicx}
\usepackage{color}
\usepackage{here}
\usepackage{amsfonts}

\begin{document}

\title{{\LARGE On the propagation of social epidemics in social networks
under S\@.I\@.R\@. model.}}

\author{ {\large {\bf Horacio Castellini}} }
\email{hcaste@ifir.edu.ar}
\affiliation{{\rm Dpto\@. de F\'{\i}sica, F\@.C\@.E\@.I\@.A\@.,
Pellegini 250,
2000 Rosario}}

\author{ {\large {\bf Lilia Romanelli}} }
\email{lili@ungs.edu.ar}
\affiliation{{\rm Instituto de Ciencias, Universidad Nacional de General
Sarmiento, J\@.M\@. Gutierrez 1150, 1613 Los Polvorines}}

\begin{abstract}
The S\@.I\@.R\@. model (Susceptible, Infected, Recovered or Died) was proposed
by chemistry Willam Kermack (1927) and the mathematician G. Mc\@. Kendrick 
(1932). the model supposes to divide to the individuals of a population in
three categories. Susceptible to be infected, Infected and Recovered (immune 
or died by the disease). On the other hand has been a similarity in 
the evolution of epidemics of infect aerial, the computer science 
propagation of virus and the propagation of social paradigms(fashion, rumor,
etc\@.) it calls modernly ``Social Epidemics''. In this work it is tried to use
this model in different types from social networks, real or not. 
In order to evaluate an meta-analysis of results
allows to investigate under wich conditions the topology of the network is 
excellent and that networks are equivalent. 
The result obtained can have relevance in study of propagation of 
epidemics of different types.
\end{abstract}

\maketitle

\section{Introduction}
The use of mathematical models to study the evolution of epidemics is 
not a newness\cite{r1}. At the moment it is known enough on behavior of 
transmission of diseases according to its characteristics like 
knowing the limits and the possibilities these models. 
The S.I.R. model developed by Scottish chemistry 
William Ogilvy Kernack (1927)\cite{r2}
and mathematician A.G. Mc Kendrick (1932)\cite{r3}
was one on the first attempts.
Unfortunately the articles fell in the forgetfulness until 1979 which 
a well-known article of Anderson and May in the Nature magazine\cite{r4}
turned east model in the departure point of the present studies. 
On the other hand, model S.I.R. generates a description in three phases 
of the course of an epidemic: starting (of slow growth), explosive and 
remission. Implicitly it is assumed that the contacts between the members 
of a population are purely random. This can even give relatively fit 
results for many diseases in virus transmission in networks of computers. 
Where it would be enough that it is considered to the users like susceptible 
individuals to the disease. This is thus because in general the electronic 
mail or the air (in the case of influenza) are opened enviroment that 
contacts to all with all with relative promiscuity then the randomness 
hypothesis can work enough\cite{r5}. 
In this work different types from real social networks obtained by real or 
fictitious study ({\em Scale-Free} {\bf SF})\cite{r6}
and {\em Small World} {\bf SW})\cite{r7}. 
Where a population of survivors was considered soon to carry out 
Meta Analysis of the data and thus to verify under which conditions 
the topology of the network is excellent and to investigate 
if the fictitious networks are equivalent to the real ones.

\subsection{Theoretical bases}
A network is a set of relations (bonds or edges) between a defined series 
of elements (nodes, vertices or actors). 
Formally a network is a graph defined as the triada $G = (V,E,\gamma)$ 
where {\em V} is the set of vertices, {\em E} is the set of edges and 
$g :E \to V$ so that $g(e) = \{v, w\}$. I\@.E\@., {\em g(\@.)} assigns to each 
edge a pair of vertices. 
Nowadays to facilitate the study of the networks the binary matrices 
instead of the whole matrices are used. Then an isomorphism $f:G \to B_n$ 
can be defined, where $B_n$ is a binary and symmetrical matrix of well-known 
dimension {\bf nxn} this matrix is called adjacency matrix (AM). 
The sociologists take like convention that in the rows is located the 
actors (exits or egos), however in the columns are located the attributes 
or related actors (entered or alter), this convention is used in this work. 
As it is possible to be appreciated the AM contains all sensible information 
on the network reason why it allows to elaborate algorithms in MA without 
considering to the network. 
Another form to characterize to the network is through the well-known 
histogram of rows also named {\em prestige of an actor} who in the case 
of the graphs agrees with the histogram of columns or 
{\em popularity of an actor}. 
At the moment the connection probability is used, $P(k)$. $P(k)$ is the 
probability that a chosen vertex at random has $k$ edges. 
According to the functional form of the tail of this histogram, 
$k \to \infty$, the type of network can be classified in: 
{\bf exponential}, when $P(k)\approx e^{-\lambda k}$; 
{\bf scale-free}, when $P(k)\approx k^{-2-\gamma}$ with $\gamma>0$; 
{\bf broad-scale} when he is scale-free with a steep cut; 
and {\bf single-scale} when it has a fast decay.

\section{An Ad-Hoc algorithm of epidemic}
The algorithm of epidemic developed in this work supposes to know AM. 
This matrix can be obtained from investigations in a closed community, 
is to say does not interact with other communities, like as a 
list of electronic mail or an isolated population. Or of an opened community, 
it is to say interacts with with other communities, as it is a system of 
the news in Internet or countries connected by massive means of transport 
of people. To each actor has a structure of data with three 
following properties is assigned:

\begin{enumerate}
\item {\bf Susceptibility or threshold (U)}: It is the minimum value 
from which the actor interchanges information with its pairs. 
\item {\bf Internal state}: It is a state represented by a 
binary vector of 32 bits.
\item {\bf Health}: It can have three possible states, alive or 
susceptible (state S), ill (state E) and dead (state M).
\end{enumerate}

Of course when the actor dies automatically they eliminate the connections 
that connect with a the network. Unlike model S.I.R., in this work one 
assumes that an actor who becomes ill can cure itself but he does not 
stop being susceptible to return to become ill. 

On the other hand the edges of the graph are characterized by the 
following properties:

\begin{enumerate}
\item {\bf State}: It is a binary value that single taking two values 
``0'' to indicate that there is not connection between actor "i" with 
actor "j". Or ``1'' to indicate that a relation between both exists.
\item {\bf Infecting}: It is an assigned value of random form and 
uniforms in the rank $C \in [0,1]$. This it indicates the probability 
of infect between actors and it is not the same one for each pair of actors. 
\end{enumerate}

The algorithm consists of interacting all the population of connected actors 
(this is defined as epoc) considering that each pair of related actors 
interacts when in a Bernoulli raffle with probability given by value 
$C$ be favorable. Soon the condition of {\bf following proximity} is 
verified. Two actors are next or neighboring if the distance of Hamming 
between its vectors of state is minor that the minimum of its thresholds. 
Then if this condition is verified two actors can interact. 
In the following listing the pseudocode shown the algorithm of sharing 
of information between the proximity actors $A_1$ and $A_2$.

\begin{verbatim}
Share Algorithm
 for each actor nondead I and each actor nondead J do
  U<- mínimun(threshold(I), threshold(J));
  if Hamming_distance(I,J) < U then
    for each bit do
     q<- Bernoulli raffle with probability C(I,J)
     if q>0 then J.bit <- I.bit
     else q <- Bernoulli raffle with probability C(I,J)
      if q>0 then I.bit <- J.bit
      end if
    end if
   end for
 end if
end for
end Share
\end{verbatim}

As it is possible to be appreciated the sharing of information is 
probabilistic and not deterministic as the case of the opinion 
models\cite{r8}. 
However if the actors are not verified the  following proximity condition  
the operation does not take place and their vectors of state do not change. 
This criterion is of agreed with the sharing of information 
codified in a data channel. 

The criterion of health adopted for an actor $A$ is related to the value 
of the norm of Hamming of its internal state. It is to say an actor 
it remains in state S if the norm of Hamming of its internal state, $H(A)$, 
is greater than the threshold than this has assigned, $U_A$. However if it 
happened the opposite the actor turn to the state E and if after the next 
interaction it happened that $H(A)<U_A$ the actor turn to state M soon 
to disappear of the network assigning the value ``0'' to all their 
possible connections. Of another form the actor recovers his state S. 

In this model one takes like control parameter the common threshold value $U$ 
to all the actors and the number of initial actors $N$. 
On the other hand, initially the network consists of a population of $K$ 
infected actors (that is to say, actors who have the state E as initial 
condition). In addition the location to infected actors is uniform 
in all the network to guarantee a random state in the population.

\section{Methods and results}
Four types of networks, two networks obtained in experimental works 
and two toy networks were used. the first toy model used it was a network 
Small-World type ({\bf SM}) with N=4096 actors with probability of 
random reconexión of 10\%. The second toy model was a network 
Scale-Free ({\bf SF}) with N=4096 actors with $\gamma=0.1$, this value 
is agreed with the experimental observations in real networks. 
Real social networks were obtained from two types of different societies 
in Internet. A society corresponds to a closed society as it happens in 
the lists of electronic mails ({\bf LM}) and the other society 
corresponds to opened societies as it happens in news groups  
and the forums ({\bf DM}). 
In the empirical networks they are not connected for that reason each 
experimental network was reduced to its connected maximum component.

In all the cases one test of 512 experiment by network of 2000 times where 
the threshold varying between $10<U<20$. 
The used statistical estimators were the absolute frequency of survival, 
the absolute frequency of final ill, the absolute frequency of infected 
initial. Except for the last one, all these variates with probability 
distribution obtained empirically from the tests. In Table I the probability 
of survival based on the threshold for the different types from networks 
can be appreciated.

\begin{center}
Table I: Survival Probability \\
\begin{tabular}{|c|c|c|c|c|}
\hline
Threshold & SM & SF & LM & DM \\
\hline
10 & 0.989 & 0.989 & 0.989 & 0.99 \\
11 & 0.973 & 0.965 & 0.967 & 0.971 \\
12 & 0.93 & 0.883 & 0.929 & 0.907 \\
13 & 0.827 & 0.758 & 0.891 & 0.791 \\
14 & 0.679 & 0.641 & 0.821 & 0.674 \\
15 & 0.549 & 0.572 & 0.702 & 0.586 \\
16 & 0.406 & 0.448 & 0.563 & 0.485 \\
17 & 0.298 & 0.35 & 0.246 & 0.231 \\
18 & 0.192 & 0.241 & 0.246 & 0.231 \\
19 & 0.118 & 0.173 & 0.179 & 0.15 \\
20 & 0.07 & 0.096 & 0.097 & 0.07 \\
\hline
\end{tabular}
\end{center}

In most of the networks for $U=15$ the survival probability it is $P(S)>50$\%. 
On the other hand the variance of survivors (Table II), 
V(S), shown that is increased 
until reaching a maximum value soon to diminish. When one studied the 
final networks of survivors this increase of the variance was due to 
the great dispersion of final graphs is to say: by each successive 
test the final population of survivors was totally different.

\begin{center}
Table II: Variance of Survival \\
\begin{tabular}{|c|c|c|c|c|}
\hline
Threshold & SM & SF & LM & DM \\
\hline
10 & 2.23 & 2.53 & 3.01 & 2.15 \\
11 & 5.37 & 18.07 & 25.94 & 9.65 \\
12 & 29.37 & 181.47 & 132.2 & 89.21 \\
13 & 160.84 & 661,06 & 425,17 & 363.69 \\
14 & 498.03 & 1224.94 & 1634.64 & 808.81 \\
15 & 665.51 & 1769.68 & 3704.39 & 1444.53 \\
16 & 744.71 & 1920.67 & 6018.13 & 2023.32 \\
17 & 876.2 & 1906.55 & 6992.23 & 2180.64 \\
18 & 910.31 & 1684.03 & 5590.49 & 1657.92 \\
19 & 849.35 & 1194.13 & 4565.42 & 1301.69 \\
20 & 574.98 & 700.03 & 2713.59 & 441.93 \\
\hline
\end{tabular}
\end{center}

Soon it was applied meta-analysis between networks of survivors for values of 
U=15 by two reasons. {\em First}: probability P(S) is greater than 50\% 
what allows to have a social structure not so destroyed. 
{\em Second}: the V(S) is great this allows to guarantee a 
great variety of final societies.

In the following tables it is the comparative results in each particular 
situation. DM and LM correspone to the analysis group of  which compares 
them with the control group given by networks SM and SF. 
The dead and alive numerical values correspond to the mean value of each test. 
The variable Q represents the sum in a column and the variable R the sum 
in a row.

\begin{center}
Table III Netowrk Meta-analysis of LM vs SM \\
\begin{tabular}{|c|c|c|c|}
\hline
 & LM & SM & R \\
\hline
Dead & 60 & 2398 & 2558 \\
Alive  & 140 & 1698 & 1838 \\
\hline
Q & 200 & 4096 & \\
\hline
\end{tabular}
\end{center}

\begin{center}
Table IV Netowrk Meta-analysis of DM vs SM \\
\begin{tabular}{|c|c|c|c|}
\hline
 & DM & SM & R \\
\hline
Dead & 83 & 2398 & 2481 \\
Alive  & 117 & 1698 & 1815 \\
\hline
Q & 200 & 4096 & \\
\hline
\end{tabular}
\end{center}

\begin{center}
Table V Netowrk Meta-analysis of LM vs SF \\
\begin{tabular}{|c|c|c|c|}
\hline
 & LM & SF & R \\
\hline
Dead & 60 & 2465 & 2525 \\
Alive  & 140 & 1631 & 1771 \\
\hline
Q & 200 & 4096 & \\
\hline
\end{tabular}
\end{center}

\begin{center}
Table III Netowrk Meta-analysis of DM vs SF \\
\begin{tabular}{|c|c|c|c|}
\hline
 & DM & SF & R \\
\hline
Dead & 83 & 2465 & 2548 \\
Alive  & 117 & 1631 & 1748 \\
\hline
Q & 200 & 4096 & \\
\hline
\end{tabular}
\end{center}

\section{Conclusions and result analysis}
In a prospective analysis of data where the probability of death 
in one given network {\bf X} is: 
\begin{equation}
p_X=P(M/X)=\frac{\texttt{Dead}}{\texttt{Q}_X}
\end{equation}
After from previous tables $p_{SM}=0.58$; $p_{LM}=0.30$; $p_{DM}= 0.41$ 
and $p_{SF}=0.60$, then if it is defined as null hypothesis, $H_0$, 
to not exist difference in probability between two networks. 
I\@.E\@. if $H_0$ is accepted soon is fulfilled a 95\% of certainty 
that $p_1=p_2$ or $DR=p_1-p_2=0$ and both networks are equivalents. $DR$ 
is known as difference of proportions whose variance comes given by:
\begin{equation}
V(DR)=\frac{p_1 \, (1-P_1)}{Q_1}+\frac{p_2 \, (1-p_2)}{Q_2}
\end{equation}
here it is assumed that the random varibles have statistical independence. 
Soon from the interval of following confidence:
\begin{equation}
I=DR\pm 1.96 \sqrt{V(DR)}
\end{equation}
If $DR=0$ belongs at this interval it accepts in probability $H_0$ 
other wise it is rejected and it posible to conclude:

\begin{enumerate}
\item When it was compared networks LM and DM were $I=0.02\pm0.09$. 
In this cituación it is clear that $H_0$ cannot be rejected and both 
networks are equivalent.
\item When it was compared networks SF and SM were $I=0.02\pm0.021$. 
Marginally it is possible to be accepted that both networks are equivalent, 
but to affection of the analyses they can be taken like equivalent.
\item When it was compared LM with SM were $I=0.28\pm0.06$. 
That clearly shown that there is to reject the null hypothesis.
\end{enumerate}

As a conclusion it is possible to be appreciated that there is no 
equivalence between the theoretical networks and the obtained ones 
by experimental work when the proposed algorithm is applied to them. 
It is to say in this prospective analysis the exposed actors to a network 
type DM or LM they have more probability of life then the behavior is diffent
on surviving population of the found one in toy networks. 
This fact brings like consequence the nonreliability of the simulations 
on propagation of aerial epidemics of virus in theoretical, 
letting a door opened to find networks theoretical more realists.
% \section{References}


\begin{thebibliography}{99}
\bibitem{r1} Pascal Cr\'epey, Fabi\'an Alvarez and Marc Barth\'elemy,
{\em Epidemic variability in complex networks} arXiv:cond-mat/0602562,
{\bf 2006}.
\bibitem{r2} W\@. O\@. Kermack and A\@. G\@. McKendrick
{\em A contribution to the mathematical theory of epidemics} Proceeding
of the Royal Society of London, Vol\@. 115, Issue 772, 700-721, {\bf 1927}.
\bibitem{r3} W\@. O\@. Kermack and A\@. G\@. McKendrick
{\em A contribution to the mathematical theory of epidemics. The
problem of Endemicity} Proceeding of the Royal Society of London, 
Vol\@. 138, Issue 834, 55-83, {\bf 1932}.
\bibitem{r4} Anderson RM and May RM
{\em Population biology of infectious diseases: Part I.} 
Nature Aug 2;280(5721):361-7 {\bf 1979}.
\bibitem{r5} M\@. Newman, S\@. Forrest and J\@. Balthrop
{\em Email networks and the spread of computer viruses}, 
Phys\@. Rev\@. E 66, 035101 {\bf 2002}.
\bibitem{r6} A\@. Barabasi and A\@. Reka
{\em Emergence of sacaling in random networks} Science 286 pp 509 {\bf 1999}.
\bibitem{r7} D\@.J\@. Watts and S\@.H\@. Strogatz
{\em Collective dynamics of small-world networks} Nature 393 pp 440-442
{\bf 1998}.
\bibitem{r8} M\@.F\@. Laguna G\@. Abramson and D\@.Zanette
{\em Vector opinion dynamics in a model of social influence}
Physica A, vol. 329, p. 459-472, {\bf 2003}.
\end{thebibliography}
\end{document}